\documentclass[useAMS,usenatbib]{mn2e}

\title[Chemical composition of NGC2419]
  {{News from the Galactic suburbia: the chemical composition of the remote globular cluster 
  NGC 2419}
  \thanks{Based on data obtained at the W. M. Keck Observatory, which is operated as a scientific 
  partnership among the California Institute of Technology, the University of California, and the 
  National Aeronautics and Space Administration. The Observatory was made possible by the generous 
  financial support of the W. M. Keck Foundation. }
  }
   
\author[Mucciarelli et al.]
  {A. Mucciarelli,$^1$ M. Bellazzini,$^2$ R. Ibata,$^3$ T. Merle,$^4$ S. C. Chapman,$^5$ 
   \newauthor
  E. Dalessandro,$^1$   A. Sollima$^6$
  \\
  $^1$ Dipartimento di Astronomia, Universit\`a 
  degli Studi di Bologna, Via Ranzani, 1 - 40127, 
  Bologna, Italy
  \\
  $^2$ INAF-Osservatorio Astronomico di Bologna, Via Ranzani, 1 - 40127, 
  Bologna, Italy
  \\
  $^3$Observatoire Astronomique, Universit\'e de Strasbourg, CNRS, 11, 
  rue de l'Universit\'e, F-67000 Strasbourg, France
  \\
  $^4$Universit\'{e} de Nice Sophia-antipolis, CNRS (UMR 7293), Observatoire de la C\^{o}te d'Azur, 
  Laboratoire Lagrange, BP 4229, 06304 Nice, France
  \\
  $^5$Institute of Astronomy, Madingley Road, Cambridge CB3 0HA, United Kingdom
  \\
   $^6$INAF- Osservatorio Astronomico di Padova,
   Vicolo dell'Osservatorio 5, I--35122 Padova, Italy
  }

\usepackage{graphicx}

\pagerange{\pageref{firstpage}--\pageref{lastpage}} \pubyear{2002}

\def\LaTeX{L\kern-.36em\raise.3ex\hbox{a}\kern-.15em
    T\kern-.1667em\lower.7ex\hbox{E}\kern-.125emX}

\begin{document}

\label{firstpage}

\maketitle

\begin{abstract}

We present the chemical analysis of 49 giant stars of the globular cluster NGC~2419, 
using medium resolution spectra collected with the multi-object spectrograph DEIMOS@Keck.
Previous analysis of this cluster revealed a large dispersion in the line strength of the 
infrared Ca~II triplet, suggesting an intrinsic star-to-star scatter in its Fe or Ca 
content. From our analysis, we assess that all the investigated stars share the same 
[Fe/H], [Ca/Fe] and [Ti/Fe] abundance ratios, while a large spread in Mg and K abundances is detected. 
The distribution of [Mg/Fe] is bimodal, with $\sim$40\% of the observed targets 
having subsolar [Mg/Fe], down to [Mg/Fe]$\sim$--1 dex, a level of Mg-deficiency never observed before in globular clusters.
It is found that the large dispersion in Mg abundances is likely the main origin of the observed dispersion of the Ca~II triplet 
lines strengths (that can be erroneously interpreted in terms of Fe or Ca abundance scatter) because Mg plays a relevant role in 
the atmosphere of giant stars as an electron donor. A strong depletion in the Mg abundance leads to an increase of the line strength 
of the Ca~II triplet, due to the variation in the electronic pressure, at a constant Fe and Ca abundance.
Finally, we detect an anti-correlation between Mg and K abundances, not easily explainable within 
the framework of the current nucleosynthesis models.

\end{abstract}

\begin{keywords}
stars: abundances -- stars: atmospheres -- stars: evolution -- stars: Population II -- 
(Galaxy:) globular clusters: individual (NGC~2419)
\end{keywords}

\section{Introduction}

The old and metal-poor cluster NGC~2419 is by far the most luminous globular cluster (GC) residing 
in the outermost fringes of the Milky Way (MW) halo, $\ga 10$ times brighter than any other cluster 
having $R_{GC}\ge 40$~kpc, \citep[see][for discussion and comparison with remote clusters in the halo of M31]{gal07}. 
This unusual feature, coupled with a half-light radius much larger than that of ordinary globulars of similar luminosity 
led several authors to the hypothesis that NGC~2419 can be the remnant of an originally larger system, like a nucleated 
dwarf galaxy \citep{mvdb,ucd}.

Since the kinematics and the stellar mass function of NGC~2419 are fully consistent with a typical GC made up of stars and 
stellar remnants \citep[see][and references therein]{iba11a,micml}, the key test to establish the actual nature of the system 
is to search for inhomogeneities in the abundance of chemical elements heavier than Aluminium among member stars. A spread in 
the heavy elements up to the iron-peak group would indicate that the system was a site of chemical evolution driven by Supernovae (SNe), 
implying that, at the epoch of SNe explosions, the progenitor of NGC~2419 was sufficiently massive to retain their highly energetic ejecta 
\citep[$>$few$\times 10^6~M_{\sun}$;][]{baum}, hence it was likely a dwarf galaxy. On the other hand,
GCs are observed to display virtually no spread in heavy elements and large (and correlated) spreads in light-elements 
(Na, O, Mg, Al, in particular), likely associated with a spread in He abundance \citep[see][and references therein]{grat12}. 
These signatures are generally believed to trace early chemical evolution driven by much less energetic polluters than SNe, 
like massive Asymptotic Giant Branch stars \citep[AGB][]{dant02,dercole08} or Fast Rotating Massive Stars 
\citep[FRMS, see][and references therein]{dec07a,dec07b}, whose ejecta can be retained in systems in the plausible range of mass of 
proto-GCs. The anti-correlation between Na and O abundances is so ubiquitous in GCs (and non-existent in the field) that 
it has been proposed as a fundamental defining feature for globulars 
\citep[with respect to open clusters and galaxies][]{carretta_def}. 

The large distance of NGC~2419 \citep[$\sim$91 kpc][]{dicriscienzo11} has prevented the detailed spectroscopic analysis of a large sample 
of its stars, that is needed to perform this chemical test, until a couple of years ago\footnote{\citet{shetrone01} provided abundance analysis 
from high resolution spectroscopy for just {\em one} star of NGC~2419.}. 
Recently \citet[][C10 hereafter]{cohen10} presented the analysis of medium-resolution Keck-DEIMOS spectra around the Calcium triplet 
(near 8600 \AA, CaT) for 43 Red Giant Branch (RGB) stars of NGC~2419. Using other GCs of known [Fe/H] as calibrators and adopting a 
constant [Ca/Fe] value they translated the measured pseudo- Equivalent Width of CaT and the magnitude difference between the stars and the 
cluster Horizontal Branch (HB, V-V$_{HB}$) into [Ca/H] values, as is usually done for [Fe/H] \citep[see][and references therein]{rut,bat,starkenburg}. 
C10 found a significant spread in the [Ca/H] values derived with this method ([Ca/H]$_{CaT}$ hereafter): the distribution showed a strong 
peak at [Ca/H]$_{CaT}\simeq -1.95$ and a long tail reaching [Ca/H]$_{CaT}\simeq -1.45$. This relatively large spread in a heavy element 
\citep[Fe or Ca, see][]{starkenburg} was interpreted by C10 as providing additional support to the hypothesis that NGC~2419 ``...is the remnant 
of a dwarf galaxy accreted long ago by the Milky Way...''. As a consistency check C10 obtained Fe and Ca abundances by spectral synthesis 
of Fe and Ca lines (other than CaT) that are present in their spectra. The puzzling result was that, while the Ca abundances 
are in reasonable agreement with those from CaT, no significant spread in Iron abundance was detected. 

\begin{figure}
\includegraphics[width=84mm]{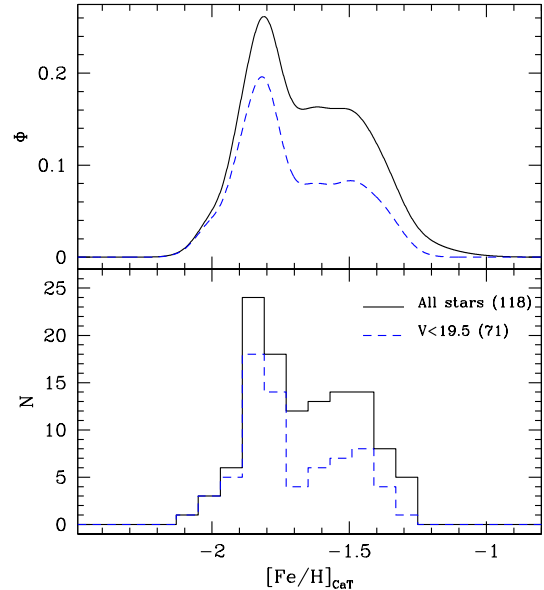}
\caption{Metallicity distribution derived from CaT, presented as ordinary histograms (lower panel) or as generalised histograms 
\citep[a representation that removes the effects due to the choice of the starting point and of the bin width, see][]{laird}. 
Note that the distribution does not change if only the brightest stars are considered (dashed lines).} 
\label{hisfecat}
\end{figure}

\begin{figure}
\includegraphics[width=84mm]{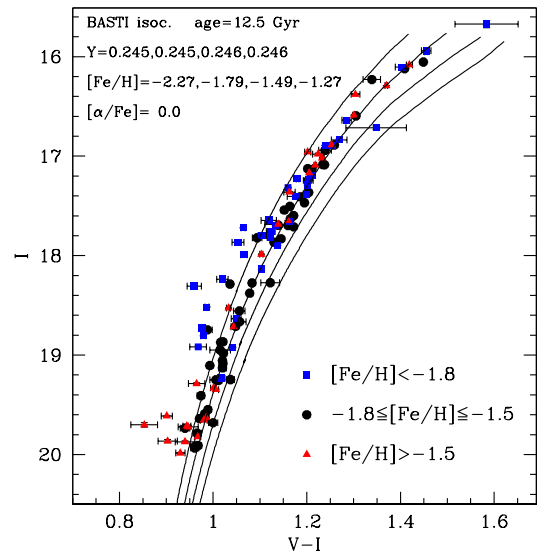}
\caption{Colour-Magnitude Diagram for the 118 stars included in the metallicity distribution from 
CaT shown in Fig.~\ref{hisfecat} from the photometric catalog assembled by \citet{b07} (see Section~\ref{CaT}).
Different symbols are adopted for different metallicity ranges.A grid of isochrones of different 
metallicity from the BaSTI set is overplotted \citep{pietr04}, for reference.} 
\label{cmd_CaT}
\end{figure}

Available photometry is not of much help in settling this issue, since in this low metallicity regime the sensitivity of optical colours to 
variations of metallicity is pretty weak. Moreover, even if a tiny colour spread is observed along the RGB, a spread in metallicity is not
necessarily the only viable explanation. \citet{dicriscienzo11b} used exquisite HST photometry to show that a small but significant colour 
spread is indeed detected at the base of the RGB, however this can be fully accounted for by the Helium spread that they invoke to reproduce
the complex HB morphology of the cluster, within the classical scenario of multiple populations in GCs \citep{grat12}.

Later \citet[][C11 hereafter]{cohen11} used high-resolution Keck-HIRES spectra to derive the detailed abundance of several chemical elements 
in seven bright RGB members of NGC~2419. Interestingly, (a) the stars studied by C11 do not display any spread in [Fe/H] and [Ca/Fe] in excess 
to what is expected from the observational uncertainties, and (b) the abundance pattern of these stars is very similar to what is observed in the classical  
metal-poor GC  M30 (NGC~7099), residing in the inner halo of the MW ($D\simeq 8$~kpc, $R_{GC}\simeq 7$~kpc). One of the stars in the C11 
sample (S1131), which is similar to the other six in all other aspects, was found to be extremely Mg-deficient ([Mg/Fe]=$-0.47$) and K-rich 
([K/Fe]=$+1.13$), a strong anomaly never reported before, at least in Pop-II stars (see C11 for a detailed discussion). Hence, at present, 
the results of the chemical test on the nature of NGC~2419 are still not conclusive.

Here we report on the chemical analysis of the medium-resolution Keck-DEIMOS spectra obtained by \citet[][I11a hereafter]{iba11a} and used, 
in that paper, for a thorough study of the kinematics of NGC~2419 \citep[see also][]{iba11b}. 

\section{Observations}

In this paper we analyse spectra collected with the DEIMOS multi-object slit spectrograph mounted on the Keck~II telescope. 
This dataset includes spectra of stars located along the RGB and observed with the high-resolution 1200 line/mm grating coupled 
with the OG550 filter: the spectra cover the spectral range of 6500-9000 \AA\ with a spectral resolution of $\sim$1.2 
\AA. Target selection, observations and the reduction procedure are described and discussed in full detail in I11a.

\subsection{Metallicity from CaT}
\label{CaT}
 
[Fe/H]$_{CaT}$ was derived by I11a for the stars observed with DEIMOS\footnote{I11a included in their sample also stars from the 
sample by \citet{baum09}, for which they took only the radial velocity estimate obtained by the same authors from HIRES spectra. } 
from the pseudo-EW of the three CaT lines (suitably combined) and V-V$_{HB}$ (see I11a for details). Here we slightly revise these 
estimates by (a) adopting the most recent and reliable value for V$_{HB}$, i.e. V$_{HB}=20.31\pm 0.01$ from \citet{dicriscienzo11} 
instead of $V_{HB}=20.45$ from \citet{harris}, as done in I11a, and (b) fully propagating all the observational errors into a final 
error in [Fe/H]$_{CaT}$ for each individual star. It is worth noting that there is only a tiny difference between the new [Fe/H]$_{CaT}$ 
and those published in I11a: the new [Fe/H]$_{CaT}$ are larger than the older ones by 0.037 dex on average, with a r.m.s. of just 0.0004 dex. 

In I11a [Fe/H]$_{CaT}$ was used as an additional parameter to clean the kinematic sample from interlopers. While it was noted that there 
was a significant spread in [Fe/H]$_{CaT}$, this result was neither quantitatively assessed nor discussed in detail. In Fig.~\ref{hisfecat} 
we present the metallicity distribution from CaT for the 118 stars from the most conservative sample of radial velocity members (their {\em Sample~A}) 
having valid [Fe/H]$_{CaT}$ in I11a. The distribution is roughly similar to that shown by C10: there is a strong peak at [Fe/H]$_{CaT}\simeq -1.8$ 
and an extended tail (with a possible shoulder at [Fe/H]$_{CaT}\simeq -1.5$, not seen in the distribution by C10) extending up to
[Fe/H]$_{CaT}\simeq -1.4$. Any subtle difference in the metallicity zero-point and in the shape of the distribution between this work and 
C10 can be ascribed to differences in the adopted calibration and in the sample size (118, or 71 stars, vs. 43). Hence we consider the 
agreement of our result with that of C10 as satisfying.

In the following, to quantitatively estimate the mean and the intrinsic spread ($\sigma$) of a distribution in some elemental ratio 
[X/Y] (like, e.g. [Fe/H] or [Ca/Fe]) we adopt an algorithm that explores a grid of the ([X/Y],$\sigma$) space and searches for the 
couple of parameters that maximises the maximum likelihood (ML) function defined as
\begin{equation}
ML([X/Y],\sigma)=\sum_{i=1}^N{{\frac{1}{\sigma_{tot}}}{\rm exp}{\left[-{\frac{1}{2}}{\left({\frac{[X/Y]-[X/Y]_i}{\sigma_{tot}}}\right)^2}\right]}}
\end{equation}
similar to what is currently done to estimate the mean velocity and velocity dispersion in kinematic samples \citep[see, e.g.][]{martin,walker}. 
$N$ is the number of stars in the sample and $\sigma_{tot}=\sqrt{\sigma^2+[X/Y]_{err,i}^2}$, where [X/Y]$_i$ is the abundance of the individual 
star and [X/Y]$_{err,i}$ is its associated error. The advantage of this approach is that it automatically takes into account the observational 
errors, thus providing estimates of the {\em intrinsic spread} with the associated {\em uncertainty} \citep[derived following][]{pm93}, a 
quantity that is especially valuable in the present contest.
Although the adopted ML function is well suited for deriving the parameters of Gaussian distributions, it is clearly not a good approximation for the 
distribution shown in Fig.~\ref{hisfecat} (but it turns out to be quite appropriate for the subsequent applications, see Sect.~\ref{synth}). 
Nevertheless, it can provide a useful quantitative estimate of the intrinsic spread, that is $\sigma=0.17\pm0.01$~dex, similar to the 
massive globular M54 that resides at the centre of the disrupting Sagittarius dwarf galaxy 
\citep{b08,carretta_m54pap}. A KMM test\footnote{The test compares the likelihood of the data for a single gaussian model and a double gaussian
model and provides the confidence level at which the single gaussian model can be rejected, see \citet{kmm}.} 
\citep{kmm}, performed both on the whole sample or on the sub-sample of the brightest stars ($V<19.5$), finds that a bimodal model 
is preferred to a unimodal model at the $\ge 99.9$\% confidence level, and the relative fraction of stars in the two populations 
is approximately 0.6 and 0.4. 

However, the Colour-Magnitude Diagram (CMD) of the target stars shown in Fig.~\ref{cmd_CaT} reveals that there are several stars that have
{\em virtually the same magnitude and color} but that differ by more than 0.3~dex in [Fe/H]$_{CaT}$. The adopted photometry 
is from the catalog assembled by \citet{b07} from various sources; however B, V, I  magnitudes for most of the target stars (and  
all the stars we analyse with spectral synthesis, below) are taken from the accurate absolute photometry by \citet{stet}.
To further investigate this apparent inconsistency between [Fe/H]$_{CaT}$ and photometry, and having in mind the results by C10 and C11, 
we decided to perform a full spectral synthesis analysis on a subsample of stars of the I11a sample with high Signal-to-Noise Ratio (SNR). 
This analysis is described and discussed in the following sections.

\section{Chemical analysis}
\label{synth}
\begin{figure}
\includegraphics[width=84mm]{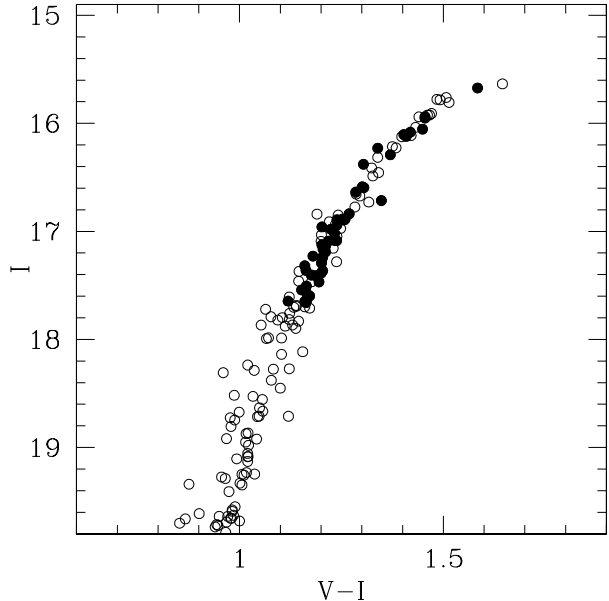}
\caption{Color-Magnitude Diagram in the I-(V-I) plane of NGC 2419  with the spectroscopic targets analysed here marked as black points.
The photometric catalog is that assembled by \citet{b07}, in particular the source of the photometry for the 
spectroscopic targets is \citet{stet}.}
\label{cmd}
\end{figure}

Among the stars recognised as cluster members by I11a in their {\sl Sample A}, we selected stars with sufficient SNR 
(typically SNR$>$40 per pixel) to perform a reliable chemical analysis based on spectral synthesis. All the stars have V-band 
magnitude $<$18.8 and their position in the CMD is shown in Fig.~\ref{cmd}. Even if the sample is clearly dominated by RGB 
stars, a few Asymptotic Giant Branch (AGB) stars are likely included. This is not expected to have any effect on the chemical 
analysis performed below. 

\subsection{Atmospheric parameters}

The atmospheric parameters ($T_{eff}$, surface gravities and microturbulent velocities) of the target 
stars cannot be derived directly from the spectra (due to the low spectral resolution and the small 
number of available lines) and we estimated them from the photometry.
Temperatures are computed by employing the empirical $(V-I)_0$-$T_{eff}$ transformation by 
\citet{alonso99} based on the InfraRed Flux Method. The dereddened color $(V-I)_0$ is obtained assuming 
for all the stars a color excess of E(B-V)=~0.08$\pm$0.01 mag \citep{dicriscienzo11} and 
the exctinction law by \citet{mccall04}. 
The \citet{alonso99} relation for the $(V-I)_0$ colour is computed in the Johnson photometric system, 
thus, our I Cousin magnitudes have been converted into I Johnson magnitudes by following the prescriptions 
by \citet{bessell79}. 
Internal errors in $T_{eff}$ due to the uncertainties in photometric data and reddening are of the order 
of about 30-50 K.

The surface gravities are computed through the Stefan-Boltzmann relation, assuming the photometric $T_{eff}$, 
and a typical mass of 0.75 $M_{\odot}$, according to an isochrone from the BaSTI dataset \citep{pietr04} with 
age of 13 Gyr, Z=0.0003 and alpha-enhanced ([$\alpha$/Fe]=~+0.4 dex) abundances patterns (see C10 and C11). 
For the luminosities, we used the dereddened V-band magnitudes with the bolometric corrections from \citet{alonso99} 
and the distance modulus of 19.71$\pm$0.08 mag \citep{dicriscienzo11}.
The gravities can be determined within $\pm$0.05 dex by propagating the uncertainties in $T_{eff}$, evolutive mass 
(of the order of 0.05 $M_{\odot}$), distance modulus and photometry.

Finally, the microturbulent velocities are calculated by adopting the relation between $v_t$ and log~g 
derived by \citet{kirby09}: this relation provides values of $v_t$ of $\sim$1.8-2.0 km/s. 
Formally, the uncertainties in log~g translate into a typical error in $v_t$ of 0.01 km/sec. Because the majority 
of the lines analysed here are located in the saturated/damped portion of the curve of growth, these lines are often 
sensitive to the photospheric velocity fields and we adopted a conservative (and more realistic) 
error of $\sim$0.2 km/s.
The adopted atmospheric parameters for all the investigated targets are listed in Table~1.

\subsection{Model atmospheres and linelist}

We began the analysis by employing one-dimensional, LTE, plane-parallel geometry model atmospheres computed 
with the code ATLAS9 (originally developed by R. L. Kurucz) that treats the line opacity through the Opacity 
Distribution Functions (ODFs) method. 
The ATLAS9 models are computed by using the ODFs calculated by \citet{castelli04} with [M/H]=--2 dex, 
$\alpha$-enhanced chemical patterns and without the inclusion of the approximate overshooting in the 
calculation of the convective flux. For some stars with abundance patterns that 
differ heavily from those adopted in the ODFs computation, specific model atmospheres are computed 
with the code ATLAS12 that employs the opacity sampling method and allows one to calculate 
model atmospheres with arbitrary chemical composition
\citep[see][for a detailed description of the handling of opacity by this code]{castelli05}. 

Adopting the photometric atmospheric parameters and a guessed metallicity value of 
[M/H]=--2 dex, we calculated synthetic spectra along the entire wavelength range of the 
observed spectra, including all the atomic and molecular transitions available in the last version of 
the Kurucz/Castelli linelist \footnote{http://wwwuser.oat.ts.astro.it/castelli/linelists.html}.
All the synthetic spectra used for the analysis have been generated with the code SYNTHE in its Linux 
version \citep{sbordone04,sbordone05} and convoluted
with a gaussian profile to mimic the spectral resolution of the DEIMOS spectra.

We selected absorption lines that are unblended at the DEIMOS resolution and with 
the atmospheric parameters and metallicity of our targets.
We included in our linelist 19 Fe~I lines, 5 Ti~I lines, the Mg~I line at 8806.7 \AA, 
the CaT line at 8542 and 8662 \AA\ and the resonance K~I line at 7699 \AA\ 
(the other K~I transition at 7665 \AA\ being heavily blended with telluric absorption). 
Reference Solar values are from \citet{gs98}.

\subsection{Abundance analysis}

The traditional derivation of abundances through the measured equivalent widths can be difficult 
in low resolution spectra, due to the line blanketing conditions that make the continuum location 
quite complex. Thus, we resort to the comparison with synthetic spectra including all the molecular 
and atomic transitions.
Each selected line has been analysed independently by performing a $\chi^2$-minimisation between the 
normalised spectrum and a grid of synthetic spectra within a spectral window centred around the line 
(in the case of DEIMOS spectra the typical width of the fitting window is $\sim$4-5 \AA). 
The synthetic spectra were rebinned at the pixel-step (0.33 \AA/pixel) of the observations. 
A preliminary normalisation of all the spectra is performed by fitting the entire spectrum 
with a Chebyshev polynomial.
For each line, we adjusted locally the normalisation  along a region of $\sim$60 \AA\ 
around the line. When a first abundance is determined from $\chi^2$ minimisation, the spectrum is 
divided by the best-fit synthetic spectrum and the quotient spectra is fitted with a spline. 
The resulting spline is used to normalise again the observed spectrum and a new $\chi^2$ minimisation 
is performed \citep[the same approach has been adopted also by][]{shetrone09,kirby11}.

To derive the Ca abundance we employed a slightly different methodology, by fitting not the entire line 
profile but only the wings of the two strongest components of the CaT, at 8552 and 8662 \AA. 
The CaT line at 8498 \AA\ has been excluded because its wings are marginally sensitive to the 
abundance, at variance to the other two lines that are located in the damped part of 
the curve of growth and they are more sensitive to abundance with respect to saturated lines.
We excluded the core region of these features because {\sl (i)}~it is heavily saturated, 
{\sl (ii)}~it suffers from relevant departures from LTE, and {\sl (iii)}~its shape depends 
on the thermal stratification of the model atmosphere in the outermost atmospheric layers 
and basically the canonical one-dimensional model atmospheres are not able 
to describe well the thermal structure of the upper layers where the line core forms.

The K resonance line at 7699 \AA\ suffers from departures from LTE conditions. Unfortunately, 
large grids of corrections as a function of the atmospheric parameters are not available for this transition. 
\citet{takeda09} and \citet{andri10} provided abundances of [K/Fe] computed by including the departures from LTE 
for globular clusters and extreme metal-poor halo stars. The magnitude of their corrections agrees very well 
and, for the resonance line used in our work, the correction is quite small \citep[see Fig.~6 in][]{takeda09}. Because of the small 
range of metallicity and $T_{eff}$ of our targets, we decide to apply a unique value (--0.30 dex) of the NLTE correction 
for all the targets. 

Concerning the CaT lines discussed here, NLTE corrections affect only the core of the lines 
(as pointed out by \citealt{spite}, see their Fig.~1, but see also \citealt{starkenburg} and references therein), 
while their wings are not affected by the influence of the NLTE and can be used to derive 
reliable Ca abundances under the LTE assumption. \citet{spite} derived 
Ca abundances of metal-poor stars by fitting the wings of the CaT lines 
with both LTE and NLTE profiles, finding that the effect of NLTE on the 
wings of these lines is negligible (less than 0.1 dex).

Uncertainties in the fitting procedure and in the atmospheric parameters were taken into account 
and added in quadrature (see Table 1). The uncertainty due to the fitting procedure is computed by resorting 
to MonteCarlo simulations: for each spectral line, Poissonian noise is injected in the best-fit synthetic 
spectrum (after the re-mapping at the pixel-scale of the observed spectrum) and re-analysed as described above.
One hundred MonteCarlo events are computed for each line and the dispersion of the derived abundance distribution is taken to be the 1$\sigma$ uncertainty. Typically, the uncertainty in the fitting procedure ranges from $\pm$0.09 dex for SNR=~100 
to $\pm$0.25 dex for SNR=~40. Only for the Ca abundances, the effect of the photon noise is quite small
($\pm$0.02 dex for SNR=~100 and $\pm$0.05 dex for SNR=~40), because of the large number of pixels used in the $\chi^2$ 
minimisation. 

For Fe, Ca and Ti (for which we have measures from different lines), we computed  
a weighted average abundance for each star by using the error from the MonteCarlo simulations as a weight; 
for the typical internal uncertainty we assumed the dispersion of the mean normalised to the root mean 
square of the number of used lines. For K and Mg (for which one only line is measured) we take
the 1$\sigma$ level of the MonteCarlo distributions as the internal error for each line.

\section{Chemical homogeneity and inhomogeneity in NGC 2419}

The derived relative abundances of Fe, Mg, Ti, Ca and K for the target stars are reported in Table~1, together with 
the associated uncertainties. The first issue we want to focus on is the lack of spread in Fe and Ti abundance in our data. 
Adopting the ML algorithm we find
$\langle{\rm [Fe/H]}\rangle=-2.09\pm0.01$ with $\sigma_{\rm [Fe/H]}=0.00\pm 0.03$, and 
$\langle{\rm [Ti/Fe]}\rangle=+0.29\pm0.02$ with $\sigma_{\rm [Ti/Fe]}=0.00\pm 0.04$.
{\em The agreement with the results by C11 is excellent} as, with the same algorithm, we obtain 
$\langle{\rm [Fe/H]}\rangle=-2.08\pm0.05$ with $\sigma_{\rm [Fe/H]}=0.00\pm 0.08$, and 
$\langle{\rm [Ti/Fe]}\rangle=+0.28\pm0.05$ with $\sigma_{\rm [Ti/Fe]}=0.00\pm 0.07$, from their sample of seven 
stars\footnote{We have considered here the abundances derived by C11 from TiII lines. From TiI $\langle{\rm [Ti/Fe]}\rangle=+0.12\pm0.02$ 
with $\sigma_{\rm [Ti/Fe]}=0.00\pm 0.07$}.

A small intrinsic spread in [Ca/Fe] is detected, $\langle{\rm [Ca/Fe]}\rangle=+0.46\pm0.01$ with $\sigma_{\rm [Ca/Fe]}=0.09\pm 0.01$, 
while from C11 we get $\langle{\rm [Ca/Fe]}\rangle=+0.14\pm0.06$ with $\sigma_{\rm [Ca/Fe]}=0.00\pm 0.07$. The non-null spread in [Ca/Fe] 
found in our data may be partially due to an underestimate of the errors on the Ca abundance we derive from the wings of the CaT lines, 
since these lines are in a different part of the curve of growth with respect to all the other lines considered here. 
Correlation between uncertainties in 
Fe and Ca abundances may also play a role, as [Ca/Fe] appears to correlate with [Fe/H] even if the spread in [Fe/H] is not statistically 
significant, while the correlation between [Ca/H] and [Fe/H] is much weaker, if any. This view is supported by the fact that we do not 
detect an intrinsic spread in [Ca/H]: $\langle{\rm [Ca/H]}\rangle=-1.63\pm0.01$ with $\sigma_{\rm [Ca/H]}=0.00\pm 0.02$. We conclude that the 
spread in Calcium abundance is absent or very small.

Contrary to Fe, Ca and Ti, large star-to-star spreads in the Mg and K content 
are detected among the stars of NGC~2419: $\langle{\rm [Mg/Fe]}\rangle=+0.05\pm0.08$ with $\sigma_{\rm [Mg/Fe]}=0.56\pm 0.06$ and 
$\langle{\rm [K/Fe]}\rangle=+0.92\pm0.08$ with $\sigma_{\rm [K/Fe]}=+0.51\pm 0.06$. As we shall see in detail below, a single gaussian 
model in not appropriate for the observed [Mg/Fe] and [K/Fe] distribution, since these are clearly bimodal. The average and intrinsic 
spread for [Mg/Fe] are in reasonable agreement with the C11 results ($\langle{\rm [Mg/Fe]}\rangle=+0.29\pm0.13$ with 
$\sigma_{\rm [Mg/Fe]}=0.30\pm 0.10$; see Sect.~\ref{mgdep}, below).

C11 provides K abundance only for the anomalous star S1131. The discussion of the K abundances is deferred to Sect.~\ref{mgk}.

The star Iba11$\_$83 was also observed by C11 (star S1209 in their sample): 
our $T_{\rm eff}$ is lower than that derived by C11 by 200 K and this difference arises mainly from 
the different colour-$T_{\rm eff}$ transformations and colour excess employed in the two studies. 
We found that the differences with respect to C11 are basically ascribable to the difference in $T_{\rm eff}$; only for Ca 
is the discrepancy large. Taking into account the corresponding iron abundances, our Ca abundance is higher than that found by C11 
by $\sim$0.3 dex, and this difference is only partially explainable with the employed $T_{\rm eff}$ scales.
We can suppose that the residual difference is due to the adopted Ca lines.

\subsection{Metallicity from CaT in Mg-deficient atmosphere}
\label{CaTMg}

The large intrinsic spreads in [Ca/H]$_{CaT}$ and in [Fe/H]$_{CaT}$ reported by C10 and I11a, respectively, and found again in 
Sect.~\ref{CaT}, above, requires an additional discussion, since our spectral synthesis analysis, in fact, confirmed the results by C11, 
i.e. there is no intrinsic scatter in Iron abundance and just a small (if any) spread in Calcium. 

The observed large variance of the CaT strength is not due to an Fe or Ca abundance spread 
but might be explained in light of the large range covered by the Mg abundance.
In fact, Mg plays a relevant role in the opacity of cool stars, being one of the most important electron 
donors, contributing to the formation of the H$^-$ ions, the main source of opacity in these stars 
together with Fe, Al and the other $\alpha$-elements (the relative contribution of each element depends 
on the temperature, gravity and optical depth).
A depletion of Mg implies a lower number of free electrons available to form H$^-$ ions, leading to a decrease 
of the  H$^-$ opacity. The strength of the CaT lines increases, decreasing the electron pressure \citep[as pointed out by][]{shetrone09}.
Thus, a large depletion of the Mg abundance produces an increase of the strength of the CaT lines at fixed Ca and Fe abundance.

\begin{figure}
\includegraphics[width=84mm]{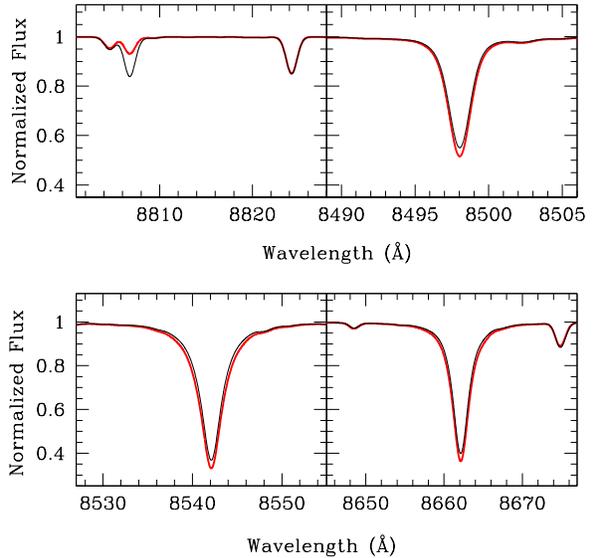}
\caption{Comparison between the synthetic spectra around the Mg~I lines at 8806.7 \AA\ 
(upper-left panel) and the three CaT lines 
computed by employing ATLAS12 model atmospheres with $T_{eff}$=4200 K, 
log~g=0.6, [M/H]=--2 dex, $\alpha$-enhanced patterns, but with [Mg/Fe]=+0.4 dex 
(black spectrum) and [Mg/Fe]=--1.0 dex (red spectrum). The synthetic spectra are at the 
same spectral resolution as the DEIMOS spectra.}
\label{synt}
\end{figure}

\begin{figure}
\includegraphics[width=84mm]{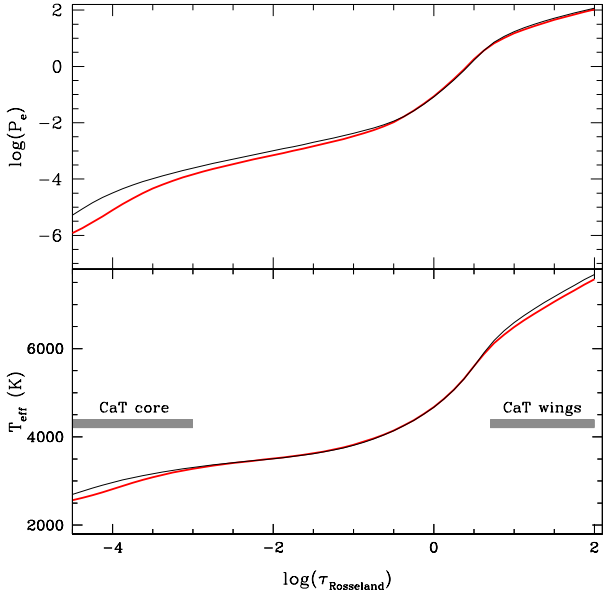}
\caption{Comparison between the two ATLAS12 models 
described in the caption of Fig.~\ref{synt}. Upper panel: electronic pressure as a function 
of the Rosseland optical depth. Lower panel: temperature as a function of the Rosseland 
optical depth. The grey regions indicate the regions of line formation for the core and the 
wings of the Ca~II triplet}.
\label{model}
\end{figure}

Fig.~\ref{synt} shows the comparison of synthetic spectra around the Mg~I line (first panel) and the three CaT lines 
computed by adopting ATLAS~12 model atmospheres with $T_{eff}$=~4200 K, log~g=~0.6 and [M/H]=--2 dex, and with [Mg/Fe]=+0.4 dex 
and [Mg/Fe]=--1.0 dex (black and red line respectively). In the latter case, the strength of the CaT lines significantly increases 
with respect to the synthetic with [Mg/Fe]=+0.4 dex (note that the lines of elements other than Mg and Ca, that are also visible in the spectra, 
are identical in the two cases). We stress that the two synthetic spectra are identical in terms of overall metallicity, individual abundance 
patterns (in particular the same Fe and Ca abundances)  and atmospheric parameter, and the only difference is in the Mg abundance.
Fig.~\ref{model} illustrates the behaviour of the electronic pressure and of the temperature as 
a function of the Rosseland opacity for the two ATLAS~12 model atmospheres, where the differences 
in terms of $T_{eff}$ and $P_e$ are appreciable (in particular in the outermost layers where the core 
of the CaT lines forms and in the deepest regions where the wings of CaT form).

\begin{figure}
\includegraphics[width=84mm]{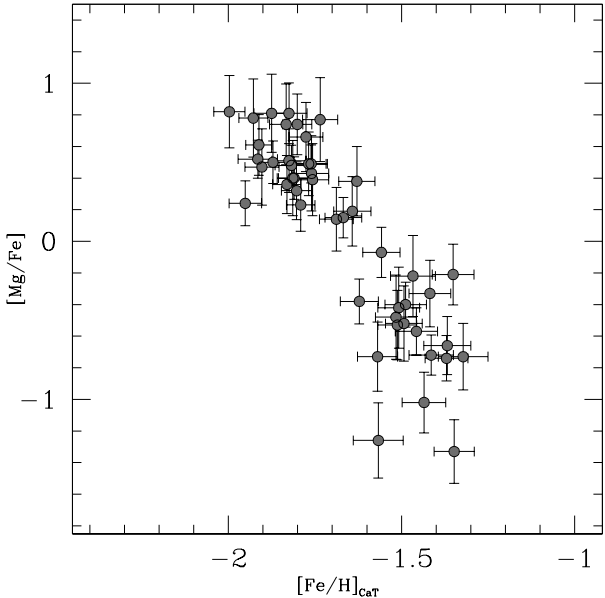}
\caption{Behaviour of the [Mg/Fe] abundance ratio as a function of the metallicity derived 
from the CaT lines. The [Mg/Fe] abundances do not include NLTE corrections: as discussed 
in Section~\ref{mgdep}, the NLTE corrections are always negative (thus lowering the [Mg/Fe] ratios) and 
their magnitudes are a function of the Mg abundance: the most Mg-rich stars can be overestimated by 0.2-0.3 dex, while 
the most Mg-poor stars are basically unaffected by departures from NLTE.}
\label{mgfe}
\end{figure}

As a confirmation of this effect, we detect a clear anti-correlation between our [Mg/Fe] abundance 
ratio and [Fe/H]$_{CaT}$  (Fig.~\ref{mgfe}). The iron-rich stars (with [Fe/H]$_{CaT} >$--1.7) dex have 
[Mg/Fe]$<$0 dex, and the bimodality in [Fe/H]$_{CaT}$ noted in Sect.~\ref{CaT} reflects the bimodal distribution 
of Mg abundance (see below).

\begin{figure}
\includegraphics[width=84mm]{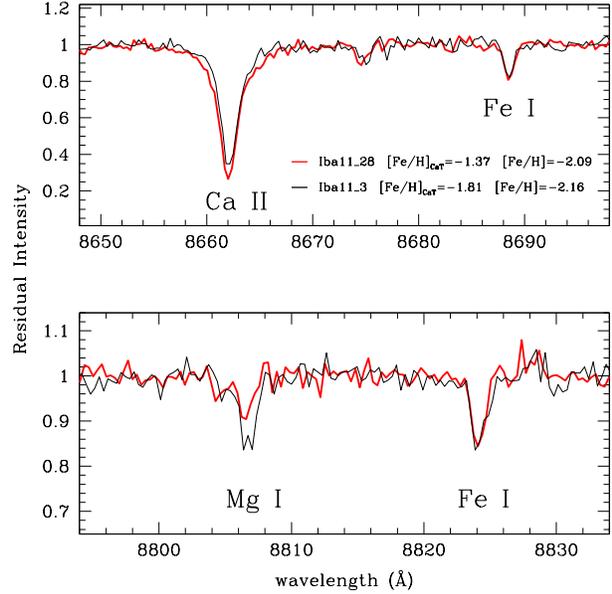}
\caption{Comparison between the spectra of the stars 
Iba11$\_$28 (red line) and Iba11$\_$3 (black line) around the CaT line at 8662 \AA\ 
(upper panel) and around the Mg~I line at 8806.7 \AA. Note that the two spectra 
have very similar atmospheric parameters but different [Fe/H] based on the CaT lines.
In particular they have $T_{eff}=4201~^oK$ and $T_{eff}=4224~^oK$, log~g=0.60 and  log~g=0.62, respectively, and the same value of 
microturbulent velocity, $v_t=1.99$ km~s$^{-1}$.}
\label{portion}
\end{figure}

The link between the Mg abundance and the intensity of the CaT lines is appreciable 
also in Fig.~\ref{portion} that compares portions of the spectra for the two target stars 
Iba11$\_$28 and  Iba11$\_$3 (red and black line respectively), that have very similar $T_{eff}$ and log~g 
(see Table 1) and different [Fe/H]$_{CaT}$ (by more than 0.4~dex, but indistinguishable Iron abundance as derived from Fe lines). 
The difference in the line strength 
of the CaT line at 8662 \AA\ is evident (upper panel), as well as the difference in the Mg~I line 
in the lower panel. On the other hand, the Fe~I lines visible in the spectra are very similar to each other.

Note that the Ca abundances that we derived from spectral synthesis of the wings of the CaT lines take into account 
the effect of the Mg abundance: the wings were fitted by using synthetic spectra computed 
with the corresponding Mg abundance derived from our analysis 
\footnote{For sake of completeness we repeated the analysis of all the stars with [Mg/Fe]$<$--0.2 dex 
by using ATLAS~12 model atmospheres computed with the corresponding abundances. Basically, the difference 
with respect to the analysis performed with the ATLAS~9 models is of a few hundredths of dex.}. 
When we repeated the fitting procedure  
by using only synthetic spectra with [Mg/Fe]=+0.4 dex (the value of the employed ODFs), the Ca abundance 
of the Mg-deficient stars are over-estimated by about 0.2--0.3 dex. This difference is smaller than 
the Ca abundance range found by C10 (spanning about 0.6-0.7 dex) because the measurement of the equivalent 
widths of these lines includes also the contribution of the line core, that is highly affected by the variation 
of the electronic density (see Fig.~\ref{synt}) but less sensitive to the Ca abundance. 

The Mg depletion  observed among the stars of the cluster might explain 
both the large distributions of [Ca/H]$_{CaT}$ by C10 and of [Fe/H]$_{CaT}$ by \citet{iba11a}.
As additional confirmation of this interpretation, the Mg-poor star S1131 discussed by C11 
has also one of the highest [Ca/H]$_{CaT}$ values in the whole C10 sample.


Finally, we stress that the strength of CaT lines depends on the {\em global} budget of the free electrons, thus, in addition to Mg, 
other elements that are electron donors in the atmospheres of cool stars can contribute to shape the CaT lines. In particular, the Al 
abundance deserves a note of caution, since in ordinary GCs [Al/Fe] anti-correlates with [Mg/Fe] \citep{grat12}. In principle,
an enhancement of Al abundance leads to an increase of the free electrons, that could counterbalance those lacking because of Mg depletion. 
This may be a source of concern for the interpretation of the abundance spreads derived from CaT presented above, since, unfortunately, we lack 
Al abundance for our target stars. We investigated the effect of different Al abundances (in the range $0.0\le$[Al/Fe]$\le +1.5$) coupled with 
a Mg depletion ([Mg/Fe]=--1) on the strength of CaT lines, using suitable ATLAS12 model atmospheres. 
We found that the strength of CaT decreases appreciably only for [Al/Fe]$\ge$1 dex, thus mitigating the effect of the Mg depletion only in 
this extreme regime. It is interesting to note  that C11 finds an Al abundance much lower than this for the only Mg-deficient star in their sample, 
S1131 ([Al/Fe]$=+0.45$). Assuming that this value is representative for the Al abundance in Mg-deficient stars of NGC~2419, the effect on the CaT lines 
should be negligible.

\section{MAGNESIUM depletion}
\label{mgdep}

The stars in our sample appear to belong to two groups, according to their Mg abundance (see Fig.~\ref{mgfe} and Fig.~\ref{mgkk}).
A first group is more or less in the range covered by stars in other GCs, $0.0\la {\rm [Mg/Fe]}\la +0.8$ dex \citep{carretta_uves}. 
Consequently we will dub these stars {\em Mg-normal}. The second group is made of {\em Mg-deficient} stars having 
$-1.4\la {\rm [Mg/Fe]}\la 0.0$ dex, a range usually not reached by stars in Galactic globulars (see below).  
The KMM test confirms that the [Mg/Fe] distribution is bimodal with a confidence level $>99.9$\%. A fraction of 62\% of the stars 
are attributed to the Mg-normal population, having $\langle{\rm [Mg/Fe]}\rangle=+0.47$ dex, and the remaining 38\% to the 
Mg-deficient population, having $\langle{\rm [Mg/Fe]}\rangle=-0.62$ dex.

Intrinsic dispersions in the Mg abundance are usually observed in GCs and explained in the framework of self-enrichment 
processes, occurring in the early stages of GC evolution.
The majority of the Milky Way GC stars for which Mg abundances are available covers a range from +0.2 to +0.5 dex, the only 
known cluster having Mg-deficient stars is NGC~2808, where {\em three}  RGB stars with [Mg/Fe] between --0.3 and --0.1 dex 
have been detected \citep{carretta_uves}. Note that NGC~2808 is characterised by a huge Na-O anti-correlation, with the presence 
of super O-poor stars, and (likely) a wide and multi-modal distribution in He abundances \citep{dant05,pi2808,angie}.
Also, two RGB stars with [Mg/Fe]=--0.21 and --0.31 dex have been measured in the Large Magellanic Cloud old GC NGC~1786 \citep{mucciarelli09} 
and both the stars share a very large ($<$--0.4 dex) depletion of [O/Fe].
Hence Mg-deficient stars are rare in GCs and seem to be linked to the hypothesised second  generation of stars formed from the ejecta of massive 
AGB stars \citep{dercole08} or FRMS \citep{dec07b} of the first generation, and characterised by high helium and sodium content and strong 
depletion of O and Mg \citep[see][and references therein]{grat12}.

Recent analyses of the bimodal HB morphology of NGC~2419 \citep{dalessandro,sandquist} show that a relevant fraction 
($\sim$30\%) of the HB population is composed of extremely hot stars which have been tentatively interpreted as second-generation stars 
with high helium content (Y=~0.42) by \cite{dicriscienzo11b}. The fraction of these extreme HB/Blue Hook stars is fairly similar to the 
fraction of Mg-deficient stars observed in our sample, and the postulated 
bimodality in the He content is intriguingly reminiscent of the bimodal distribution of [Mg/Fe] detected here. 

However the range of [Mg/Fe] covered by our stars is much larger than anything ever seen before in any stellar system.
Hence, while we will come back to discuss the possible physical origin of this feature in Sect.~\ref{summa}, it is worth 
discussing briefly some details of the analysis that may lead to erroneous estimates of [Mg/Fe].

\begin{figure}
\includegraphics[width=84mm]{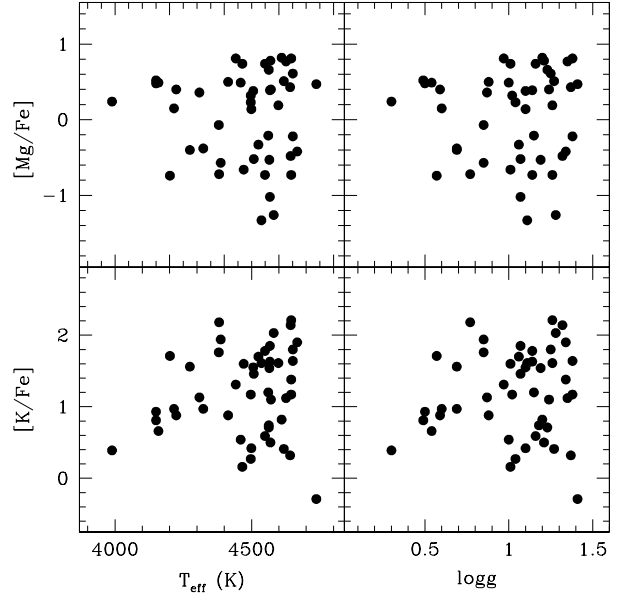}
\caption{Behaviour of [Mg/Fe] (upper panels) and [K/Fe] (lower panels) 
abundance ratios as a function of $T_{eff}$ (left panels) and log~g (right panels).}
\label{trends}
\end{figure}

In Fig.~\ref{portion}, it is clearly shown that large differences in the strength of Mg lines are
observed in the spectra of stars with the {\em same} atmospheric parameters\footnote{The considered stars differ by  
just 23~$^oK$ in the adopted $T_{eff}$, by 0.02~dex in log~g, and by 0.0~km~s$^{-1}$ in $v_t$, that is significantly less than 
the uncertainties in each parameter.}, hence some factor other than a variation in these parameters should be responsible 
of the large star-to-star differences in Mg lines. This is confirmed also by the upper panels of Fig.~\ref{trends}, showing 
that there is no discernible trend between Mg abundance and $T_{eff}$ and log~g.
Systematics in the estimates of the micro-turbulent velocity (the only atmospheric parameter that we cannot derive directly 
from photometry) does not seem able to produce spurious underestimates of the Mg abundance. In fact, the Mg lines for stars 
of similar $T_{eff}$ and log~g are distributed in both the linear and saturated portion of the curve of growth: the weakest lines 
(those with anomalous low Mg abundances) are those basically unaffected by the velocity fields. 
Changes in microturbulent velocities will change the Mg abundance for the Mg-rich stars 
but will have a small or negligible impact on the Mg-poor stars.

Furthermore, the lack of trend of [Mg/Fe] ratio as a function of $T_{eff}$ and log~g
suggests that NLTE effects play a negligible role and they are not responsible for the observed Mg range. 
As pointed out by \citet{merle11}, the NLTE corrections (in the sense $A_{NLTE}$-$A_{LTE}$) of the Mg line for a giant star 
with [M/H]=--2 dex and $\alpha$-enhanced chemical mixture are negative, hence the application of the correction would lead to even lower 
Mg abundance for Mg-deficient stars. 

In order to check the magnitude of the NLTE corrections for this Mg line, we calculated the
departures from LTE by using a modified version of the NLTE radiative transfer code MULTI \citep{carlsson86} 
and $\alpha$-poor MARCS model atmospheres \citep{gustafsson08} with the typical parameters of our targets: 
$T_{\mathrm{eff}}$=~4250~K, $\log{g}$=~1.5 and [Fe/H]=--2. We used a simplified model atom of Mg~I with 15 energy levels, 26  
radiative bound-bound transitions and 91 electron collisional bound-bound transitions \citep[the atomic data are the ones used in][]{merle11}. 
We assumed  no contribution from inelastic collision with hydrogen: this 
will produce upper limits in NLTE abundance corrections. 
The magnitude of the corrections decreases for lower Mg abundances, 
because the line is formed progressively deeper in the photosphere, reducing the contribution of the core 
(where the deviations from LTE are more relevant): for [Mg/Fe]=--1 dex the NLTE 
correction is $<$0.1 dex and the line profile is basically the same under both LTE 
and NLTE assumptions, while the maximum correction (--0.35 dex) is for high Mg abundances. 
Thus, we can exclude that the very low [Mg/Fe] values are due to NLTE effects. On the other hand, proper NLTE corrections may lower 
the Mg abundance of the most Mg-rich stars in our sample, thus leading to an upper limit in the [Mg/Fe] distribution of NGC~2419 more similar 
to that observed in other Galactic GCs of comparable metallicity (see Fig.~\ref{naneky}, below). Lacking a sufficiently fine grid of NLTE corrections 
we are forced to neglect this factor: the reader must be aware that the Mg abundance of stars having $[Mg/Fe]\ga+0.5$ can be overestimated by 0.2-0.3~dex.

As a sanity check of the Mg indicator used here, we analysed the DEIMOS spectrum of the star S1131 (kindly provided by E. Kirby 
and J. Cohen) that C11 identify as Mg-poor through the measurements of other Mg lines. When the atmospheric parameters by C11
are adopted, we derive [Mg/Fe]=--0.35$\pm$0.13, in perfect agreement with the HIRES results.
We note that our $T_{eff}$ scale is slightly hotter than that used by C11 (due to the adopted reddening value and $T_{eff}$-colour 
transformations). Still, if we derive the atmospheric parameters for S1131 following the same procedure adopted for the stars 
in our sample, as described in Section 3.1, we obtain [Mg/Fe]=--0.27$\pm$0.13, again in good agreement with C11. We stress that the [Mg/Fe] 
estimates by C11 and by us for S1131 are based on different Mg lines and the slightly higher abundance derived from the Mg 
line at 8806.7 \AA\ can be easily explained by taking into account the NLTE effects 
(note however that the two values are fully consistent within the errors, the actual difference being 0.08 dex).

Finally, we checked the impact of a strong helium enhancement and compare synthetic spectra computed 
by using ATLAS~12 model atmospheres with normal Y abundance (Y=0.24) and with Y=~+0.4. 
The net effect of a high Y value is to make the lines deeper (especially for the strong features)
but this effect is very small (typically less than 0.5\% of the line depth computed with normal Y) 
and totally negligible in the analysis of real spectra. Note that the assumption of Y=~+0.4 does 
not modify the line profile of the Mg~I line at 8806.7 \AA, thus we rule out 
that the strong depletion of Mg abundance is due to the effect of a very strong He enhancement.

Unfortunately, we cannot measure other elements involved in the chemical anti-correlations typical of GCs \citep{carretta_def}. 
O and Al cannot be detected at this spectral resolution, and the only 
Na transitions available in our spectral range are the doublet at 8183-94 \AA, 
that is heavily contaminated by telluric lines (we highlight that the 
cleaning of telluric lines at this resolution can be quite imprecise). We cannot 
assess if the Mg depletion is linked to variations of the other light elements. However, 
the very large Mg distribution of NGC~2419 could be the clue of a peculiar 
star formation history of the cluster and indicate extremely extended anti-correlation patterns.

\section{A Mg-K anticorrelation?}
\label{mgk}

\begin{figure}
\includegraphics[width=84mm]{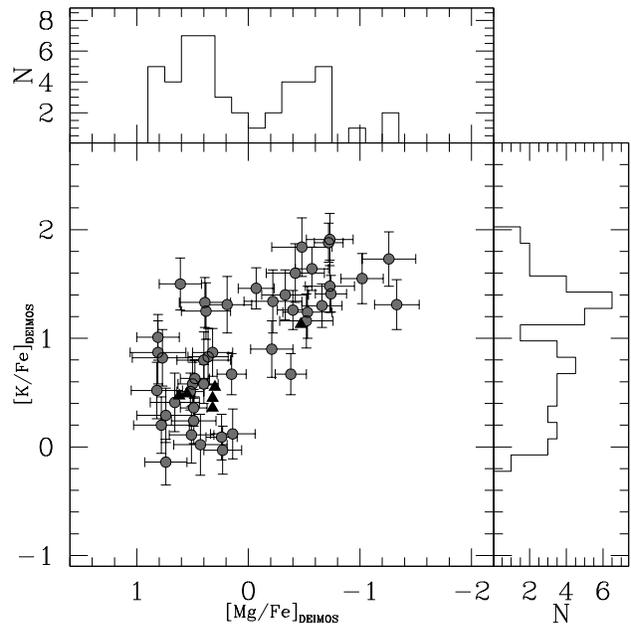}
\caption{Behaviour of [K/Fe] abundance ratios as a function of [Mg/Fe] ratios. 
Black triangles are the stars analysed by C11. The histograms of the 
[Mg/Fe] and [K/Fe] distributions are also plotted.}
\label{mgkk}
\end{figure}

Fig.~\ref{mgkk} displays the clear and significant anti-correlation between 
[Mg/Fe] and [K/Fe] that occurs in our sample.
C11 dedicated a detailed description to their star S1131, labelled 
as ``{\sl a star with a peculiar abundance pattern}", characterised by Mg depletion 
([Mg/Fe]=--0.47 dex) and K enhancement ([K/Fe]=~+1.13 dex). These abundances 
nicely match those obtained in our analysis (the stars analysed by C11
are plotted for reference in Fig.~\ref{mgkk} as black triangles).
Taking advantage of our sample of 49 stars, we can assess that this star is not an isolated 
case, but it belongs to a sub-population of stars in NGC~2419 characterised by sub-solar 
[Mg/Fe] ratios and high ($>$1 dex) [K/Fe] ratios.

We investigated some technical reasons that may be able to mimic the observed anti-correlation.
\begin{enumerate}

\item {\em Contamination by telluric lines.} The used K line is located near the red side 
of the telluric absorption A band, hence it may be prone to some contamination. 
We checked accurately for possible blending with 
telluric lines, by using as template the DEIMOS spectrum of a rapidly-rotating white dwarf. 
In all the spectra the K line is free from telluric lines. The Mg line is far from any telluric feature.
 
\item {\em Errors in the atmospheric parameters.}
Both the K and Mg abundance are virtually insensitive to the adopted gravity 
(at a level of $\pm$0.01 dex for a variation of $\delta$log~g=$\pm$0.1). 
The abundances are more sensitive to the $T_{eff}$: a variation of $\pm$50 K 
translates into a variation of the Mg number density of $\pm$0.06 dex and 
of the K number density of $\pm$0.1 dex.
The errors in $T_{eff}$ can lead a correlation between 
the Mg and K abundances but are not able to mimic a spurious anti-correlation.
In our analysis $v_t$ is not constrained spectroscopically 
(as is usually done) but we adopted an empirical relation between $v_t$ and log~g.
A variation of $\pm$0.2 km/s leads to a variation of $\mp$0.08 dex 
and $\mp$0.14 dex in Mg and K abundance respectively. Also in this case the errors 
run in the same direction and are not able to introduce 
an anti-correlation; 

\item {\em Temperature scale.} We checked the impact of a different 
$T_{eff}$ scale, assuming the temperatures inferred by projecting the position 
of each target in the CMD on the best-fit isochrone (see Section 3.1). 
The average difference between the two scales is $T_{eff}^{isochrone}$-$T_{eff}^{Alonso99}$=--97 K 
($\sigma$=~52 K). The chemical analysis with this $T_{eff}$ scale does not significantly change 
the observed dispersions both in Mg and K abundances, and does not erase the Mg-K anti-correlation.

\item {\em Departures from LTE.} We applied a unique NLTE correction to all the targets, following 
the work by \citet{takeda09}. The NLTE corrections are usually a function of the 
metallicity and the atmospheric parameters: we note that stars in our sample with the same 
parameters exhibit different K abundances, suggesting that our NLTE correction is not 
the cause of the large dispersion in the abundance of Potassium.

\end{enumerate}

The two elements are mainly produced by massive stars: K comes from hydrostatic oxygen shell burning and explosive 
oxygen burning, while Mg is produced by hydrostatic carbon burning and explosive neon burning. However, it is 
noteworthy that the current yields for K are inadequate to reproduce the observed chemical patterns in the 
Milky Way \citep[see][]{romano}. Also, Mg is destroyed through the Mg-Al cycle of the hot CNO chain in AGB stars.
However, production of  K through the proton-capture on Argon nuclei could occur in the nucleosynthesis chains 
of massive AGB stars. Unfortunately the uncertainties related to the cross-sections of these reactions prevent any reliable 
predictions; preliminary computations indicate that a simultaneous variation of the initial Argon abundance and of the 
cross-section of the proton-Argon capture can produce ejecta as enriched in K as we observe in the atmosphere of the 
Mg-deficient stars (P.Ventura and F. D'Antona, private communication).

\citet{takeda09} identify two K-rich stars in GCs M~4 and M~13, coupled 
with emission components along the H$_{\alpha}$ line profile. These findings suggest 
an increase of the velocity fields in the upper photospheric layers where the core 
of the K~I line at 7699 \AA\ forms. On the other hand, C11 ruled out 
this hypothesis, because some stars with normal K abundances exhibit H$_{\alpha}$ emissions 
larger than those observed in the K-rich star S1131. Unfortunately, the setup of our spectra 
does not include the H$_{\alpha}$ line and we cannot check this possibility. 
In any case, a rise of the turbulent velocity field in the outermost layers of the photosphere 
should affect also the core of the Mg~I at 8806.7 \AA\ in a similar fashion.

Finally, we note that the two employed transitions have very different excitational potentials 
(0 and 4.34 eV for K and Mg line respectively), thus they formed in different photospheric layers. 
In fact, the formation of low-$\chi$ lines is favoured in external layers, due to the decrease of the temperature 
with depth, while high-$\chi$ transitions occur preferentially in the deepest photospheric regions. 
In particular, the wings of the two lines form at similar depths, while the formation depth of the core 
is shifted toward low $\tau$, thus decreasing $\chi$ \citep[for comparison see Fig.~13.4 in][]{gray2}.
The K line forms in the outermost photospheric region where the thermal fluctuations 
can be relevant (in contrast to the Mg line that forms deeply). 
Also, temporal variations of the thermal structure of the photosphere cannot be excluded and they could 
introduce an additional scatter in K abundances but they do not totally explain the anti-correlation with Mg. 
In any case, the observed spread in Potassium requires an independent confirmation using other lines, as well as 
a search for the Mg-K anti-correlation in other clusters.

\section{A comment about CaT}
\label{commcat}

The CaT represents a widely used indicator of metallicity, especially for distant
stellar populations, for which high SNR, high resolution spectra are not easy to achieve. 
Basically, the combined EW of the CaT lines is a function of the metallicity and it is 
interpreted in terms of [Ca/H] or [Fe/H] \citep[see][and references therein]{starkenburg}. 

From a theoretical point of view the strength of the CaT lines is sensitive to the 
abundance of Ca (of course) and Fe, but also to the abundance of those elements that affect the 
H$^{-}$ continuum opacity, through their contribution to the electronic density.
Basically, the calibrations of the CaT  EWs are referred only to 
cases where all $\alpha$-elements abundance patterns are described by an unique value: 
for instance, stars with the same level of enhancement of each $\alpha$-element.
The [$\alpha$/Fe] value sets not only the abundance of each $\alpha$-element
but also the free electron density, which has a relevant impact in the 
continuum opacity of cool stars (as discussed above). Spurious cases, where 
the $\alpha$-elements have largely different patterns to each other, are not 
taken into account in the usual CaT  calibration, but,
as demonstrated by the case of NGC~2419, they may lead to a wrong 
interpretation of the CaT strengths. In fact, the CaT can be used as a diagnostic of the metallicity 
only for stars with {\sl canonical} chemical composition, while the method fails with 
stars with a peculiar or exotic chemical mixture, because the effective global budget of the free electrons
is not taken into account.

Mg deficiencies as extreme as those observed here seem to be very rare in nearby stellar systems (see Fig.~\ref{naneky}, 
below, and discussion in C11,  Sect.~\ref{mgdep} and Sect.~\ref{summa}). Therefore it is unlikely that their 
metallicity distributions derived from CaT are seriously biased by this effect.
However, in principle, observed dispersions of CaT in GCs (or dwarf galaxies) can be due to {\sl (i)} an 
intrinsic variation of Fe (under the assumption that all the stars share the same $\alpha$ content), 
{\sl (ii)} an intrinsic variation of Ca (under the assumption that the abundances of the electron donors do 
not vary from star to star), but also to {\sl (iii)} an intrinsic dispersion of one (or more) of those elements 
that are electron donors. Hence, in deriving metallicities of individual stars from CaT lines, it seems advisable 
to counter-check also the abundance of Mg (which should be feasible to undertake in most cases, since the Mg~I line at 8806.7 \AA\ 
lies very close to the CaT), to validate the adopted calibration and avoid possible spurious effects as those described 
in Sect.~\ref{CaTMg}.

\section{Summary}
\label{summa}

The main results derived from the analysis of 49 giant stars in NGC 2419 
are summarised as follows:\\ 

\begin{itemize}

\item All the stars share the same iron content, with an average iron abundance 
of [Fe/H]=--2.09$\pm$0.02 dex ($\sigma$=~0.11 dex), where the observed dispersion is 
fully compatible with the uncertainties. Also, [Ca/Fe] and [Ti/Fe] turn out to be homogeneous;

\item NGC 2419 exhibits a large dispersion in the Mg abundance, reaching values of [Mg/Fe]$\sim$--1 dex 
(unusual for GC stars). 
The large spread of Mg is likely the main origin of the observed dispersion of the CaT lines 
strength, previously interpreted as intrinsic dispersion in Ca or Fe. 
In fact, a Mg depletion leads to an increase of the equivalent widths of the CaT lines 
(at a constant Ca abundance). This effect is confirmed by the fact that the iron content inferred 
from the Ca triplet lines clearly anti-correlates with the Mg abundances.
However, we bear in mind that the strength of the CaT lines suffers from the abundances of all the 
elements that are electron donors and the correct line profile for these lines should be 
derived by taking into account the global budget of the free electrons;

\item The [Mg/Fe] distribution is bimodal, with about 40\% of the stars having sub-solar [Mg/Fe] abundance ratio. 
This fraction is similar to that suggested by \citet{dicriscienzo11b} 
for the extreme population stars with initial helium abundance of Y$\simeq$+0.4, according to 
the Horizontal Branch morphology of the cluster;

\item a very large spread in K content is detected among the stars of NGC 2419, spanning 
from solar values up to [K/Fe]$\sim$2 dex, with a bimodal distribution. A clearcut anti-correlation 
between [Mg/Fe] and [K/Fe] is observed, in agreement with the results already found by C11 
for the star S1131, that shows an unusual depletion of Mg coupled with a strong 
enhancement of K. 

\end{itemize}

\begin{figure}
\includegraphics[width=84mm]{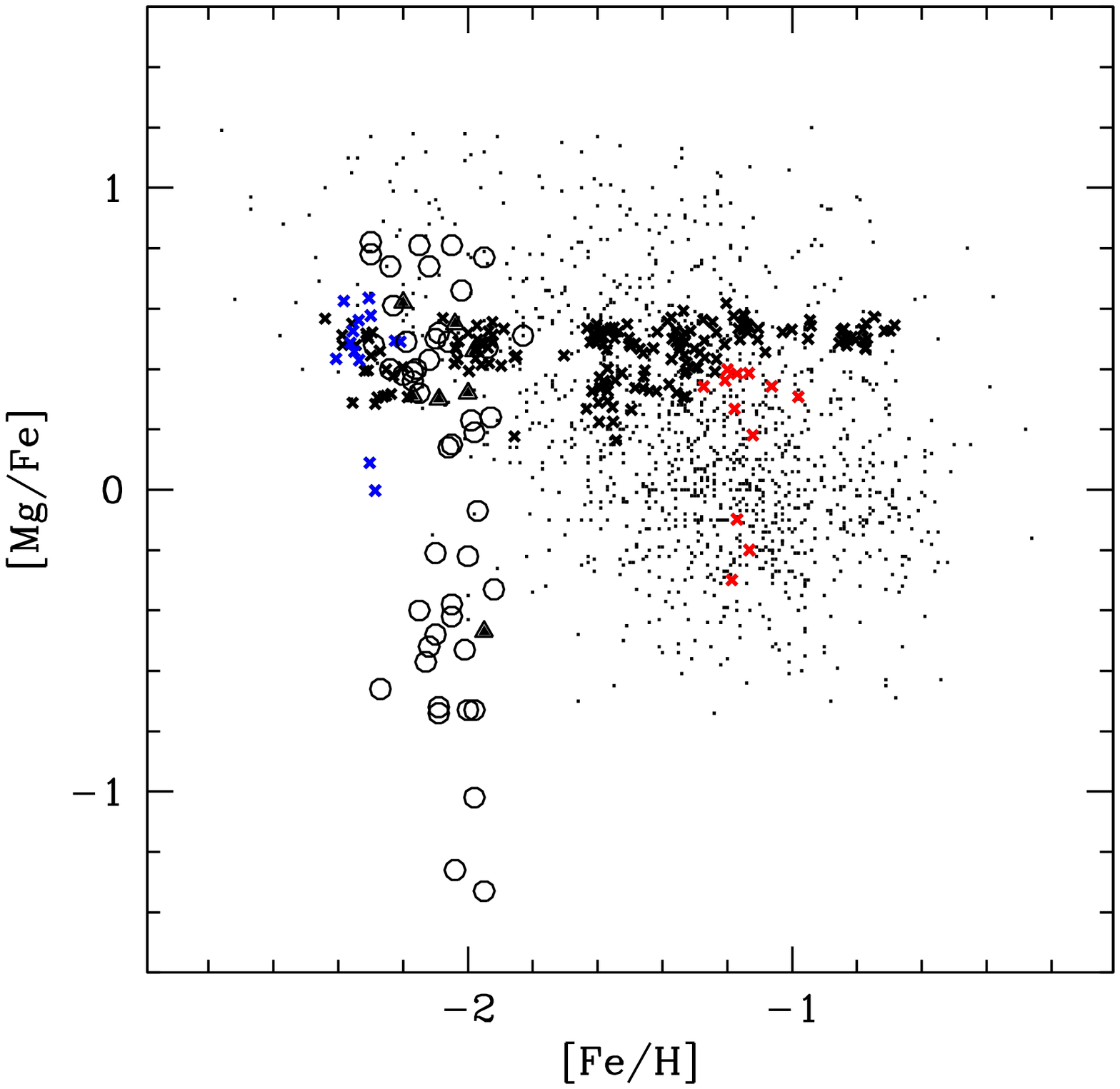}
\caption{[Mg/Fe] vs. [Fe/H] for {\em a) open circles:} stars of NGC~2419 from the present study,
{\em b) triangles:} stars of NGC~2419 from C11, 
{\em c) small dots:} stars in the dwarf spheroidal satellites of the MW from \citet{kirby11}, 
{\em d) $\times$ symbols:} stars of various Galactic GCs from \citet{carretta_uves}. Stars from NGC~7078 and 
NGC~2808 are plotted in blue and red, respectively, for reference, since they display the most Mg-deficient stars 
in the whole \citet{carretta_uves} sample.}
\label{naneky}
\end{figure}

Fig.~\ref{naneky} provides a direct illustration of the extremely unusual abundance pattern of NGC~2419. The spread 
in [Mg/Fe] is unrivalled both in GCs and in dwarf galaxies. A strong depletion of Mg is generally interpreted as due 
to a significant contribution of ejecta from SNIa to the gas from which Mg-poor stars are formed \citep[see][and 
references therein]{bekki_lmc}. Indeed, Fig.~\ref{naneky} shows that Mg-depletion with increasing metallicity is 
a common feature in dwarf spheroidal (dSph) galaxies and that [Mg/Fe]$<0.0$ stars are not rare in those systems. 
\citet{carina} reports that the two most metal-rich stars in their sample of red giants in the Carina dSph have $-0.6<$[Mg/Fe]$<-1.0$, 
not far from the most Mg-deficient stars in our NGC~2419 sample. However, the Mg depletion in dSph (as in any other 
galaxies studied until now) is always coupled with an increase of [Fe/H]\footnote{As well as a star formation history lasting 
for a few Gyrs \citep{eline}, clearly not observed in NGC~2419 \citep{dicriscienzo11b,micml}.}, since the whole effect 
is due to SNIa enriching the interstellar medium with material that is Fe-rich and poor in $\alpha$ elements, thus reducing 
the [Mg/Fe] ratio with respect to the pattern previously set up by SNII \citep[see][for a recent review]{eline}. 
This clearly does not occur in NGC~2419 where stars in the range $-1.4\la$[Mg/Fe]$\la+0.8$ are indistinguishable in terms 
of [Fe/H].

This fundamental observational fact, coupled with the analogy of the bi-modalities in [Mg/Fe] and HB morphology recalled above, 
suggest as more likely  the possibility that the observed pattern was produced by the same processes that cause the Na-O 
anti-correlations, and other signatures of the early chemical enrichment that are peculiar to GCs \citep{carretta_def,grat12}.
Further support to this hypothesis is provided by the fact that Mg-deficient stars lie systematically to the red of Mg-normal 
stars along the RGB, in the V, U-V Colour-Magnitude Diagram (Lardo et al., in preparation). In the framework of anti-correlations in GCs, 
Mg-deficient stars should correspond to Na-rich (and O-poor) stars and it is a generally observed characteristic of GCs to have 
Na-rich RGB stars that are redder than their Na-poor counterparts in CMDs that include the U-band \citep[see][for discussion and 
references]{lardo}. The colour difference is due to variations in the strength of CN and CH features at wavelengths lower than 
$\sim 4000$ \AA, that are driven by variation of C and N abundance, that, in turn, are correlated with the abundance of other 
light elements tracing the self-enrichment process in GCs \citep[like Na and O,][]{sbo11}. 

Even if the observed Mg spread can be attributed to this kind of processes, the reason for the extreme 
behaviour of this cluster, as well as the origin of the Mg-K anti-correlation, remain unclear. This latter feature, if confirmed, may shed new light 
into the whole process of GC formation, possibly providing a tool that may help to discriminate between competing models of 
self-enrichment \citep{dercole08,dec07a,dec07b}. An obvious singular characteristic of NGC~2419 is its distance from the centre of 
the Galaxy. Is it possible that a low degree of interaction with - for instance - the gas rich disc of the Milky Way has favoured 
the retention of enriched material in this cluster. \citet{micml} found some indirect evidence for the cluster having a significant 
larger total mass in the past, as postulated by \citet{dicriscienzo11b} on completely a different basis: also a large total stellar mass 
may have played a role (but this is a general feature of models of GC formation accounting for the presence of multiple populations). 
In any case, this mysterious cluster does not cease to reveal new interesting features that makes it especially worthy of further 
study.

\section*{Acknowledgments}

We warmly thank the anonymous referee for his/her suggestions in improving the paper.
We are very grateful to several colleagues for useful discussions and precious suggestions, and/or for providing their 
own data for counter-checks and comparisons:
Piercarlo Bonifacio, Angela Bragaglia, Eugenio Carretta, Judith Cohen, Franca D'Antona, Raffaele Gratton, Evan Kirby, 
Carlo Nipoti, Donatella Romano, Chris Sneden, Paolo Ventura.
M.B. acknowledge the financial support of INAF through the PRIN-INAF
2009 grant assigned to the project {\em Formation and evolution of massive star
clusters}, P.I.: R. Gratton. 
A.S. acknowledge the support of INAF through the 2010 postdoctoral fellowship grant.
R.I. gratefully acknowledges support from the Agence Nationale de la Recherche though the grant POMMME (ANR 09-BLAN-0228). 
This research has made use of NASA's Astrophysics Data System.

\begin{table*}
\begin{minipage}{160mm}
\caption{Identification numbers (from I11a), temperatures, gravities, microturbulent velocities and abundance ratios for the 
observed stars in NGC~2419.}
\begin{tabular}{lcccccccc}
\hline
ID &  $T_{eff}~[^oK]$ &  logg & $v_t$~[km s$^{-1}$] & [Fe/H]&   [Mg/Fe] & [K/Fe] & [Ti/Fe]  & [Ca/Fe] \\
\hline
         Sun        &    ---      &      ---   &   ---        &      7.50            &	  7.58           &    5.12            &	    5.02             &  6.36	      \\
\hline   
    Iba11$\_$54   &    3989	&      0.33  &   2.05	    &	  -1.93$\pm$0.13   &	0.24$\pm$0.14  &   0.09$\pm$0.21    &  0.54$\pm$0.19	   &   0.24$\pm$0.04	\\
    Iba11$\_$29   &    4150	&      0.52  &   2.01	    &	  -2.09$\pm$0.11   &	0.52$\pm$0.12  &   0.51$\pm$0.17    &  0.27$\pm$0.10	   &   0.41$\pm$0.05	\\
    Iba11$\_$83   &    4150	&      0.53  &   2.01	    &	  -2.29$\pm$0.12   &	0.48$\pm$0.13  &   0.63$\pm$0.17    &  0.25$\pm$0.13	   &   0.58$\pm$0.03	\\
    Iba11$\_$57   &    4159	&      0.57  &   2.00	    &	  -2.06$\pm$0.11   &	0.49$\pm$0.12  &   0.36$\pm$0.17    &  0.33$\pm$0.11	   &   0.44$\pm$0.03	\\
    Iba11$\_$28   &    4201	&      0.60  &   1.99	    &	  -2.09$\pm$0.11   &   -0.74$\pm$0.14  &   1.41$\pm$0.17    &  0.23$\pm$0.10	   &   0.50$\pm$0.02	\\
    Iba11$\_$3	  &    4224	&      0.62  &   1.99	    &	  -2.16$\pm$0.13   &	0.40$\pm$0.13  &   0.58$\pm$0.18    &  0.25$\pm$0.15	   &   0.53$\pm$0.02	\\
    Iba11$\_$52   &    4216	&      0.63  &   1.99	    &	  -2.05$\pm$0.11   &	0.15$\pm$0.13  &   0.67$\pm$0.19    &  0.05$\pm$0.11	   &   0.44$\pm$0.03	\\
    Iba11$\_$21   &    4323	&      0.72  &   1.96	    &	  -2.05$\pm$0.12   &   -0.38$\pm$0.14  &   0.67$\pm$0.19    &  0.21$\pm$0.14	   &   0.44$\pm$0.02	\\
    Iba11$\_$44   &    4274	&      0.72  &   1.96	    &	  -2.15$\pm$0.11   &   -0.40$\pm$0.14  &   1.26$\pm$0.18    &  0.25$\pm$0.13	   &   0.55$\pm$0.03	\\
    Iba11$\_$159  &    4381	&      0.80  &   1.95	    &	  -2.09$\pm$0.13   &   -0.72$\pm$0.13  &   1.88$\pm$0.18    &  0.36$\pm$0.12	   &   0.55$\pm$0.04	\\
    Iba11$\_$36   &    4387	&      0.88  &   1.93	    &	  -2.13$\pm$0.12   &   -0.57$\pm$0.15  &   1.64$\pm$0.20    &  0.50$\pm$0.13	   &   0.56$\pm$0.03	\\
    Iba11$\_$112  &    4380	&      0.88  &   1.93	    &	  -1.97$\pm$0.13   &   -0.07$\pm$0.16  &   1.46$\pm$0.19    &  0.25$\pm$0.10	   &   0.37$\pm$0.03	\\
    Iba11$\_$164  &    4414	&      0.91  &   1.92	    &	  -2.10$\pm$0.12   &	0.50$\pm$0.13  &   0.58$\pm$0.20    &  0.34$\pm$0.16	   &   0.43$\pm$0.05	\\
    Iba11$\_$39   &    4309	&      0.90  &   1.92	    &	  -2.17$\pm$0.16   &	0.36$\pm$0.18  &   0.83$\pm$0.26    &  0.22$\pm$0.26	   &   0.41$\pm$0.05	\\
    Iba11$\_$6	  &    4442	&      1.00  &   1.90	    &	  -2.05$\pm$0.13   &	0.81$\pm$0.19  &   1.01$\pm$0.21    &  0.32$\pm$0.21	   &   0.42$\pm$0.02	\\
    Iba11$\_$82   &    4471	&      1.04  &   1.89	    &	  -2.27$\pm$0.14   &   -0.66$\pm$0.18  &   1.30$\pm$0.20    &  0.36$\pm$0.25	   &   0.70$\pm$0.03	\\
    Iba11$\_$35   &    4460	&      1.03  &   1.89	    &	  -2.19$\pm$0.13   &	0.49$\pm$0.20  &   0.24$\pm$0.21    &  0.14$\pm$0.29	   &   0.53$\pm$0.06	\\
    Iba11$\_$51   &    4497	&      1.05  &   1.89	    &	  -2.15$\pm$0.13   &	0.32$\pm$0.18  &   0.87$\pm$0.22    &  0.25$\pm$0.16	   &   0.51$\pm$0.05	\\
    Iba11$\_$45   &    4466	&      1.04  &   1.89	    &	  -2.12$\pm$0.14   &	0.74$\pm$0.19  &  -0.14$\pm$0.21    &  0.17$\pm$0.18	   &   0.42$\pm$0.08	\\
    Iba11$\_$4	  &    4497	&      1.07  &   1.88	    &	  -1.99$\pm$0.13   &	0.23$\pm$0.17  &  -0.03$\pm$0.22    &  0.03$\pm$0.14	   &   0.33$\pm$0.03	\\
    Iba11$\_$101  &    4567	&      1.10  &   1.88	    &	  -1.98$\pm$0.14   &   -1.02$\pm$0.19  &   1.55$\pm$0.23    &  0.12$\pm$0.12	   &   0.43$\pm$0.05	\\
    Iba11$\_$7	  &    4525	&      1.09  &   1.88	    &	  -1.92$\pm$0.13   &   -0.33$\pm$0.21  &   1.40$\pm$0.23    &  0.41$\pm$0.13	   &   0.39$\pm$0.04	\\
    Iba11$\_$160  &    4508	&      1.10  &   1.88	    &	  -2.12$\pm$0.13   &   -0.52$\pm$0.24  &   1.16$\pm$0.25    &  0.34$\pm$0.26	   &   0.53$\pm$0.02	\\
    Iba11$\_$66   &    4506	&      1.13  &   1.87	    &	  -2.20$\pm$0.13   &	0.38$\pm$0.22  &   1.25$\pm$0.26    &  0.27$\pm$0.18	   &   0.58$\pm$0.05	\\
    Iba11$\_$119  &    4499	&      1.13  &   1.87	    &	  -2.06$\pm$0.15   &	0.14$\pm$0.20  &   0.12$\pm$0.23    &  0.21$\pm$0.17	   &   0.42$\pm$0.03	\\
    Iba11$\_$72   &    4536	&      1.14  &   1.87	    &	  -1.95$\pm$0.12   &   -1.33$\pm$0.20  &   1.31$\pm$0.23    &  0.29$\pm$0.16	   &   0.43$\pm$0.07	\\
    Iba11$\_$121  &    4567	&      1.17  &   1.86	    &	  -2.17$\pm$0.14   &	0.39$\pm$0.23  &   1.33$\pm$0.23    &  0.29$\pm$0.12	   &   0.53$\pm$0.06	\\
    Iba11$\_$32   &    4549	&      1.17  &   1.86	    &	  -1.98$\pm$0.13   &   -0.73$\pm$0.22  &   1.48$\pm$0.24    &  0.37$\pm$0.13	   &   0.39$\pm$0.03	\\
    Iba11$\_$43   &    4561	&      1.18  &   1.86	    &	  -2.10$\pm$0.12   &   -0.21$\pm$0.19  &   0.90$\pm$0.26    &  0.41$\pm$0.14	   &   0.59$\pm$0.02	\\
    Iba11$\_$148  &    4549	&      1.19  &   1.86	    &	  -2.24$\pm$0.14   &	0.74$\pm$0.26  &   0.29$\pm$0.25    &  0.44$\pm$0.17	   &   0.56$\pm$0.05	\\
    Iba11$\_$89   &    4610	&      1.23  &   1.85	    &	  -2.30$\pm$0.14   &	0.82$\pm$0.23  &   0.52$\pm$0.26    &  0.58$\pm$0.15	   &   0.53$\pm$0.04	\\
    Iba11$\_$96   &    4563	&      1.21  &   1.85	    &	  -2.18$\pm$0.13   &	--	       &   0.44$\pm$0.23    &  0.38$\pm$0.11	   &   0.47$\pm$0.05	\\
    Iba11$\_$11   &    4565	&      1.22  &   1.85	    &	  -2.01$\pm$0.14   &   -0.53$\pm$0.22  &   1.24$\pm$0.24    &  0.29$\pm$0.16	   &   0.41$\pm$0.03	\\
    Iba11$\_$88   &    4569	&      1.24  &   1.84	    &	  -2.30$\pm$0.12   &	0.78$\pm$0.25  &   0.20$\pm$0.26    &  0.25$\pm$0.23	   &   0.54$\pm$0.04	\\
    Iba11$\_$69   &    4651	&      1.28  &   1.84	    &	  -2.23$\pm$0.13   &	0.61$\pm$0.19  &   1.50$\pm$0.24    &  0.51$\pm$0.14	   &   0.44$\pm$0.03	\\
    Iba11$\_$10   &    4645	&      1.29  &   1.83	    &	  -2.00$\pm$0.15   &   -0.73$\pm$0.21  &   1.91$\pm$0.24    &  0.29$\pm$0.14	   &   0.53$\pm$0.05	\\
    Iba11$\_$134  &    4563	&      1.26  &   1.84	    &	  -2.02$\pm$0.15   &	0.66$\pm$0.22  &   0.41$\pm$0.27    &  0.22$\pm$0.17	   &   0.26$\pm$0.09	\\
    Iba11$\_$74   &    4571	&      1.27  &   1.84	    &	  -2.24$\pm$0.13   &	0.40$\pm$0.24  &   0.80$\pm$0.26    &  0.34$\pm$0.21	   &   0.51$\pm$0.07	\\
    Iba11$\_$1	  &    4618	&      1.30  &   1.83	    &	  -1.83$\pm$0.13   &	0.51$\pm$0.20  &   0.11$\pm$0.26    &  0.18$\pm$0.17	   &   0.14$\pm$0.08	\\
    Iba11$\_$104  &    4598	&      1.29  &   1.83	    &	  -1.98$\pm$0.14   &	0.19$\pm$0.22  &   1.31$\pm$0.26    &  0.22$\pm$0.22	   &   0.36$\pm$0.03	\\
    Iba11$\_$40   &    4581	&      1.31  &   1.83	    &	  -2.04$\pm$0.13   &   -1.26$\pm$0.24  &   1.73$\pm$0.25    &  0.28$\pm$0.13	   &   0.48$\pm$0.05	\\
    Iba11$\_$49   &    4643	&      1.35  &   1.82	    &	  -2.10$\pm$0.15   &   -0.48$\pm$0.27  &   1.84$\pm$0.27    &  0.27$\pm$0.23	   &   0.53$\pm$0.03	\\
    Iba11$\_$25   &    4667	&      1.37  &   1.81	    &	  -2.05$\pm$0.13   &   -0.42$\pm$0.26  &   1.60$\pm$0.27    &  0.05$\pm$0.16	   &   0.46$\pm$0.07	\\
    Iba11$\_$61   &    4645	&      1.37  &   1.81	    &	  -2.14$\pm$0.15   &	--	       &   1.08$\pm$0.27    &  0.47$\pm$0.18	   &   0.40$\pm$0.11	\\
    Iba11$\_$158  &    4626	&      1.38  &   1.81	    &	  -1.95$\pm$0.13   &	0.77$\pm$0.27  &   0.82$\pm$0.26    &  0.19$\pm$0.14	   &   0.33$\pm$0.06	\\
    Iba11$\_$55   &    4641	&      1.40  &   1.81	    &	  -2.12$\pm$0.13   &	0.43$\pm$0.24  &   0.02$\pm$0.28    &  0.09$\pm$0.19	   &   0.48$\pm$0.05	\\
    Iba11$\_$170  &    4737	&      1.44  &   1.80	    &	  -1.94$\pm$0.13   &	0.47$\pm$0.24  &   ---  	    &  0.38$\pm$0.14	   &   0.18$\pm$0.08	\\
    Iba11$\_$90   &    4651	&      1.41  &   1.81	    &	  -2.00$\pm$0.13   &   -0.22$\pm$0.26  &   1.34$\pm$0.29    &  0.21$\pm$0.19	   &   0.46$\pm$0.03	\\
    Iba11$\_$46   &    4645	&      1.41  &   1.81	    &	  -2.15$\pm$0.13   &	0.81$\pm$0.25  &   0.87$\pm$0.29    &  0.43$\pm$0.12	   &   0.39$\pm$0.06	\\
\hline
\end{tabular}
\end{minipage}
\end{table*}

\end{document}